# Speech Enhancement Based on Non-stationary Noise-driven Geometric Spectral Subtraction and Phase Spectrum Compensation

Md Tauhidul Islam[a], Udoy Saha[b], K.T. Shahid[b], Ahmed Bin Hussain[b], Celia Shahnaz[b,*]

[a]*Department of Electrical and Computer Engineering, Texas A&M University, College Station, Texas, USA-77840*
[b]*Department of Electrical and Electronic Engineering, Bangladesh University of Engineering and Technology, Dhaka-1000, Bangladesh*

**Abstract**

In this paper, a speech enhancement method based on noise compensation performed on short time magnitude as well phase spectra is presented. Unlike the conventional geometric approach (GA) to spectral subtraction (SS), here the noise estimate to be subtracted from the noisy speech spectrum is proposed to be determined by exploiting the low frequency regions of current frame of noisy speech rather than depending only on the initial silence frames. This approach gives the capability of tracking non-stationary noise thus resulting in a non-stationary noise-driven geometric approach of spectral subtraction for speech enhancement. The noise compensated magnitude spectrum from the GA step is then recombined with unchanged phase of noisy speech spectrum and used in phase compensation to obtain an enhanced complex spectrum, which is used to produce an enhanced speech frame. Extensive simulations are carried out using speech files available in the NOIZEUS database shows that the proposed method consistently outperforms some of the recent methods of speech enhancement when employed on the noisy speeches corrupted by street or babble noise at different levels of SNR in terms of objective measures, spectrogram analysis and formal subjective listening tests.

## 1. Introduction

Over the decades, several methods have been developed to solve the noise reduction and speech enhancement problems which are important in the area of speech processing applications, such as speech coding, speech recognition and hearing aid devices. We can divide these methods in mainly three categories based on their domains of operation, namely time domain, frequency domain and time-frequency domain. Time domain methods include the subspace approach [10, 15, 23, 35], frequency domain methods includes speech enhancement methods based on discrete cosine transform [4], spectral subtraction [3, 12, 19, 24, 34], minimum mean square error (MMSE) estimator [6, 9, 13, 25], Wiener filtering [2, 5, 28, 30] and time-frequency domain methods involve the employment of the family of wavelet or wavelet packet [1, 7, 11, 20, 32]. All these methods have their advantages and disadvantages. Time domain methods like subspace based approaches provide a trade-off between the speech distortion and residual noise. But these methods need large computation, which makes real-time processing very difficult. On the other hand, frequency domain methods provide the advantage of real-time processing with less computational load. Among frequency domain methods, the most prominent one is spectral subtraction [3, 12, 24, 34]. This method provides the facility of deducting noise from the noisy signal based on stationary nature of noise in speech signals. But this method has a major drawback of producing an artifact named musical noise, which is perceptually disturbing, made of different tones of random frequencies and has an increasing variance. The geometric approach to spectral subtraction is proposed in [26] to get rid of the musical noise in enhanced speech. In the MMSE estimator based methods [6, 9, 13, 25], the spectral amplitude of noisy signal is modified based on the minimum mean square error. A large variance as well as worst performance in highly noisy situation are the main problems of these methods. The main problem of Wiener filter based methods [2, 5, 28, 30] is the necessity of clean speech statistics for their implementations. Like MMSE estimators, wiener filters also try to reach at an optimum solution depending on the error between the computed signal and the real signal.

*Corresponding author
*Email address:* celia.shahnaz@gmail.com (Celia Shahnaz)

In the above mentioned methods, although the spectrum of a signal is a complex number, only the magnitude of the noisy speech spectrum is modified based on the estimate of the noise spectrum and phase remains unchanged. This was being done for a long time based on an assumption that human auditory system is phase-deaf, i.e., cannot differentiate change of phase, until the authors in [33] showed that the phase spectrum could also be very useful in speech enhancement. The authors used the phase spectrum in a SS based approach to obtain an enhanced speech. Later the authors in [18, 31] also justified that idea. But these methods did not consider the magnitude spectrum at all which is not suitable for most practical cases. In this paper, we propose a new noise compensation method which works on magnitude spectrum as well as phase spectrum. Noise driven geometric approach to spectral subtraction and phase spectrum compensation algorithm are used in the proposed method for obtaining the compensated magnitude and phase, respectively. A novel noise estimation technique is proposed here, which can provide a better estimate of non-stationary noise.

The paper is organized as follows. Section 2 describes the problem formulation and the proposed method. Section 3 describes results of both objective and subjective evaluation. Concluding remarks are presented in section 4.

## 2. Problem Formulation and Proposed Method

In the presence of additive noise denoted as $v[n]$, a clean speech signal $x[n]$ gets contaminated and produces noisy speech $y[n]$. The proposed method is based on the analysis modification and synthesis (AMS) framework, where speech is analyzed frame wise since it can be assumed to be quasi-stationary. The noisy speech is segmented into overlapping frames by using a sliding window. A windowed noisy speech frame can be expressed in the time domain

as

$$y^t[n] = x^t[n] + v^t[n], \tag{1}$$

where $t$ is the frame number, $t = 1, \ldots, T$, $T$ is the total number of frame. If $Y^t[\omega_k]$, $X^t[\omega_k]$ and $V^t[\omega_k]$ are the fast Fourier transform (FFT) representations of $y^t[n]$, $x^t[n]$ and $v^t[n]$, respectively, we can write

$$Y^t[\omega_k] = X^t[\omega_k] + V^t[\omega_k], \tag{2}$$

The $N$-point FFT, $Y^t[\omega_k]$ of $y^t[n]$ can be computed as

$$Y^t[\omega_k] = \sum_{n=0}^{N-1} y^t[n] e^{-\frac{j2\pi nk}{N}}. \tag{3}$$

$Y^t[\omega_k]$ is modified in the proposed method to obtain an estimate of $X^t[\omega_k]$. An overview of the proposed speech enhancement method is shown by a block diagram in Fig. 1. We can see from this figure that the magnitude spectrum of a noisy speech frame is at first modified by GA, which we denote as step-1. The modified magnitude from this step is then combined with the unchanged phase of the noisy speech spectrum. Using inverse fast Fourier transform (IFFT) and overlap and add, an intermediate speech signal is obtained. The spectrum of the intermediate speech is sent to step-2, which consists of phase spectrum compensation (PSC) [33]. PSC modifies the phase spectrum and using this phase spectrum with modified magnitude from the first step, we obtain an enhanced complex spectrum. Finally, using IFFT and overlap and add, an enhanced speech is constructed. The full AMS process is done for both steps to get full flexibilities of using different window sizes and parameters. In the following subsections, we discuss the GA and PSC methods in detail.

*2.1. Geometric Approach for Magnitude Compensation*

Geometrically, spectrum of noisy signal, $Y^t(\omega_k)$ can be represented as sum of two complex numbers $X^t(\omega_k)$ representing clean speech spectrum and $V^t(\omega_k)$ representing noise spectrum, which is shown in Fig. 2 [26]. $Y^t(\omega_k)$, $X^t(\omega_k)$ and $V^t(\omega_k)$ are the complex numbers which can be expressed in the following polar form,

$$a_Y e^{j\theta_Y} = a_X e^{j\theta_X} + a_V e^{j\theta_V}, \tag{4}$$

where $[a_Y, a_X, a_V]$ are the magnitudes and $[\theta_Y, \theta_X, \theta_V]$ are the phases of noisy, clean speech and noise spectra, respectively. Now if we express the complex numbers in a right angle $\triangle ABC$, using sine rule, we can write from Fig. 3 that

$$\begin{aligned}
\overrightarrow{AB} &= a_Y \sin(\theta_V - \theta_Y) = a_X \sin(\theta_V - \theta_X) \\
&\Rightarrow a_Y^2 \sin(\theta_V - \theta_Y) = a_X^2 \sin(\theta_V - \theta_X) \\
&\Rightarrow a_Y^2 \left[1 - \cos^2(\theta_V - \theta_Y)\right] = a_X^2 \left[1 - \cos^2(\theta_V - \theta_Y)\right] \\
&\Rightarrow a_Y^2 \left(1 - C_{YV}^2\right) = a_X^2 \left(1 - C_{XV}^2\right).
\end{aligned} \tag{5}$$

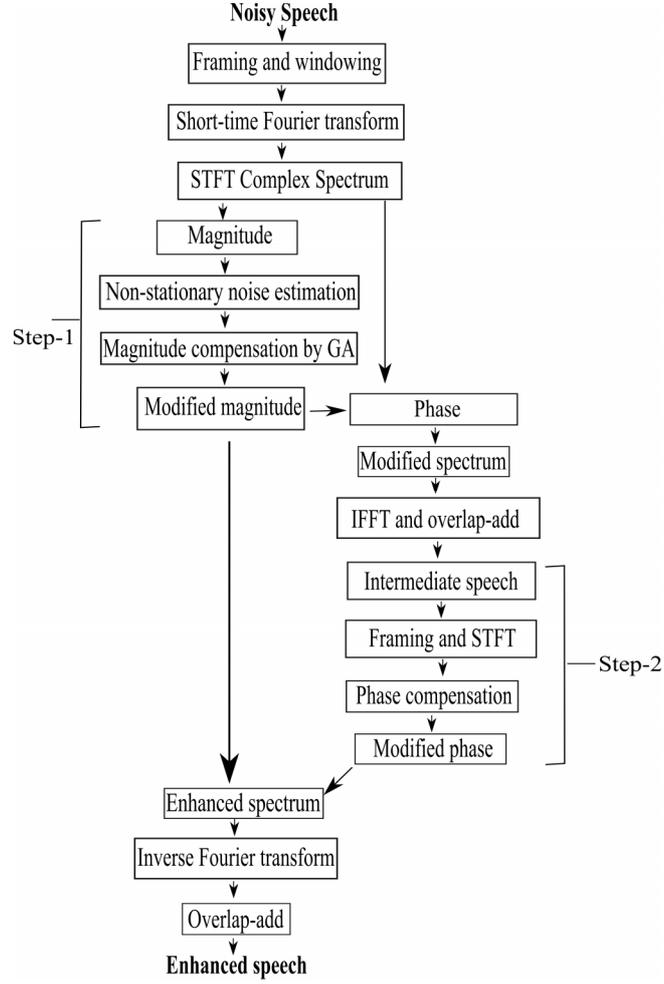

Figure 1: Block diagram of the proposed method.

The gain function of GA, $G_{GA}$ can be defined as

$$G_{GA}^2 = \frac{a_X^2}{a_Y^2} = \frac{1 - c_{YV}^2}{1 - c_{XV}^2}, \qquad (6)$$

where $c_{YV} = cos(\theta_V - \theta_Y)$ and $c_{XV} = cos(\theta_V - \theta_X)$. When $X(\omega_k)$ and $V(\omega_k)$ are orthogonal, $c_{XV}$ becomes zero. It is the case when noise and clean speech signal are uncorrelated and thus the cross-terms in eq. (5) are zero. Using cosine rules in triangle $\triangle ABC$, we can write

$$c_{YV} = \frac{a_Y^2 + a_V^2 - a_X^2}{2 a_Y a_V}, \qquad c_{XV} = \frac{a_Y^2 - a_V^2 - a_X^2}{2 a_X a_V}. \qquad (7)$$

If we divide eq. (7) by $a_V^2$ and defining $\beta = \frac{a_Y^2}{a_V^2}$, $\sigma = \frac{a_X^2}{a_V^2}$, we can write

$$c_{YV} = \frac{\frac{a_Y^2}{a_V^2} + 1 - \frac{a_X^2}{a_V^2}}{\frac{2a_Y}{a_V}} = \frac{\beta + 1 - \sigma}{2\sqrt{\beta}}, \tag{8}$$

$$c_{XV} = \frac{\frac{a_Y^2}{a_V^2} - 1 - \frac{a_X^2}{a_V^2}}{\frac{2a_X}{a_V}} = \frac{\beta - 1 - \sigma}{2\sqrt{\sigma}}. \tag{9}$$

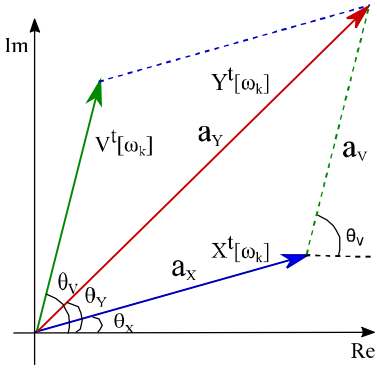
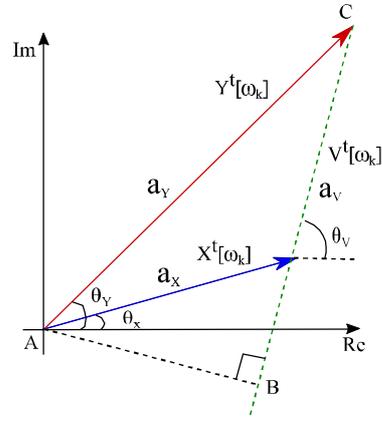

Figure 2: Representation of noisy spectrum $Y^t(\omega_k)$ in complex plane as sum of clean signal spectrum $X^t(\omega_k)$ and noise spectrum $V^t(\omega_k)$.

Figure 3: Triangle showing the geometric relationship between the phases of the noisy speech, noise and clean speech spectra.

As defined in [8], $\beta$ and $\sigma$ are the posterior and a priori SNRs respectively used in MMSE algorithm. Substituting value of $c_{YV}$ and $c_{XV}$ from eq. (8) and (9) in eq. (6), we get the gain function as

$$G_{GA} = \sqrt{\frac{1 - \left(\frac{\beta + 1 - \sigma}{4\beta}\right)^2}{1 - \left(\frac{\beta - 1 - \sigma}{4\sigma}\right)^2}}. \tag{10}$$

From the basic rule of spectral subtraction method, an estimate of clean speech magnitude spectrum can be obtained as

$$\left|Z^t[\omega_k]\right| = \left|G_{GA}[\omega_k] \cdot Y^t[\omega_k]\right|. \tag{11}$$

Aggregating the modified magnitude spectrum with the unchanged phase of noisy speech, we obtain the modified complex spectrum as

$$Z^t[\omega_k] = |Z^t[\omega_k]|e^{\angle Y^t[\omega_k]}. \tag{12}$$

After using IFFT on $Z[\omega_k]$ and overlap and add, we obtain time-domain intermediate speech $z[n]$.

*2.2. Determination of Noise*

In the proposed GA based noise reduction scheme, the noise spectrum is estimated at each silence frame as

$$|\widehat{V}^{t_s}[\omega_k]| = \begin{cases} \frac{|Y^1[\omega_k]|+\cdots+|Y^{N_s}[\omega_k]|}{N_s}, & \text{for } t = 1, \\ v_n|V^{t_s-1}[\omega_k]| + (1-v_n)|Y^t[\omega_k]|, \\ \text{otherwise,} \end{cases} \quad (13)$$

where $N_s$ is the number of initial silence frames, $v_n$ is the forgetting factor, $V^{t_s-1}[\omega_k]$ is the noise spectrum of previous silence frame and $Y^t[\omega_k]$ represents the estimated spectrum of the noisy speech at the $t$-th frame. The noise estimate at any frame $t$ can be written as

$$|\widehat{V}^t[\omega_k]| = \alpha^t|\widehat{V}^{t_s}[\omega_k]|, \quad (14)$$

where $\alpha^t$ is the tracking factor, $t_S$ refers to the index of the immediate last silence frame. Considering that this estimate of the noise spectrum is updated only during a silence period while it may change drastically with time, it is insufficient to use a constant value of the tracking factor $\alpha^t$ to compensate for the errors induced in the noise spectrum. In order to track the time variation of the noise, $\alpha^t$ should be adjusted at each frame after a silence period. According to the spectral characteristics of human speech, the low frequency band typically from 0 to 50 Hz contains no speech information. Thus, for noisy speech, the low frequency band, say $\Delta = [0, 50]$ Hz contains only noise. In view of this fact, in order to change the value of $\alpha^t$ for the $t$-th frame after a silence period we propose to use the ratio between $|Y^t[\omega_k]|$ and $|\widehat{V}^{t_s}[\omega_k]|$ in low frequency band $\Delta$ as

$$\alpha^t = \frac{\sum_\Delta |Y^t[\omega_k]|}{\sum_\Delta |\widehat{V}^{t_s}[\omega_k]|}, \quad \text{where } \Delta = [0, 50] Hz. \quad (15)$$

In the low frequency band $\Delta$ of the $t$-th frame, the variation of the noisy speech spectrum is equivalent to the noise spectrum of that frame. Thus, use of $\alpha^t$ defined in eq. 15 clearly serves as a relative weighing factor with respect to the estimated noise spectrum in eq. 13, leading to a reasonable tracking for the time variation of the noise if non-stationary.

*2.3. Phase Compensation*

If we apply STFT on $z[n]$, we obtain $Z^\tau[\omega_k]$, where $\tau$ is the frame number for step-2. In this section, the phase of $Z^\tau[\omega_k]$, which is same as the phase of the noisy speech spectrum $Y^\tau[\omega_k]$, is modified in such a way that the low energy component cancel out more than the high energy components. The modified complex spectrum by aggregating the modified phase from this step with the modified magnitude from previous step is a better representation of $X^\tau[\omega_k]$ [33].

$$\widehat{X}^\tau[\omega_k] = |Z^\tau[\omega_k]|e^{j\angle(Z^\tau[\omega_k]+\phi^\tau[\omega_k])}. \quad (16)$$

The noisy speech frame, $y^\tau[n]$ in the analysis stage is a real valued signal and therefore, its FFT is conjugate symmetric, i.e.,

$$Y^\tau[\omega_k] = Y^{*\tau}[-\omega_k]. \quad (17)$$

The conjugates can be obtained as a result of applying FFT on $y^\tau[n]$. The conjugates arise naturally from the symmetry of the magnitude spectrum and anti-symmetry of the phase spectrum. During IFFT operation as needed for clean speech synthesis, the conjugates are summed together to produce larger real valued signal. As in the previous step, we modify only the magnitude spectrum of $Y^\tau[\omega_k]$ without changing its syemmtricity, the conjugate symmetry holds for $Z^\tau[\omega_k]$. We modify the conjugates of $Z^\tau[\omega_k]$ so that they contribute constructively to the reconstruction of the clean time domain signal. For this purpose, we formulate a phase spectrum compensation function as given by

$$\Phi^\tau[\omega_k] = \Lambda \psi[\omega_k] \left| \widehat{D^\tau}[\omega_k] \right|, \tag{18}$$

where $\Lambda$ is a real-valued constant and the estimate of noise spectrum, $\widehat{D^\tau}$ is the root mean square value of $Z^\tau$, where $Z^\tau = (Z^\tau[1], \ldots Z^\tau[N])^T$ [33]. In eq. 18, $\psi[\omega_k]$ is defined as

$$\psi[\omega_k] = \begin{cases} 1, & \text{if} \quad 0 < \frac{k}{N} < \frac{1}{2}, \\ -1, & \text{if} \quad \frac{1}{2} < \frac{k}{N} < 1, \\ 0, & \text{otherwise.} \end{cases} \tag{19}$$

Here, zero weighting is assigned to the values of $k$ corresponding to the non-conjugate vectors of FFT, such as $k = 0$ and value at $k = N/2$ if $N$ is even. Since the estimate of noise magnitude spectrum is symmetric, introduction of the weighting function defined by eq. 18 produces an anti-symmetric compensation function that acts as the cause for changing the angular phase relationship in order to achieve noise cancellation during synthesis. Although discussed in [33], we will revisit the phase spectrum compensation procedure in brief in next few paragraphs for two different scenarios. For simplicity, we will denote the two complex conjugates of $Z^\tau$ as $\vec{Z}$ and $\vec{Z^*}$, of $\widehat{X^\tau}$ as $\vec{X}$ and $\vec{X^*}$ and of $\Phi^\tau$ as $\vec{\Phi}$ and $\vec{\Phi^*}$.

In Fig. 4(a), the magnitudes of the conjugates of $Z^\tau$ are considered larger than those of $\Phi^\tau$. Column one of Fig. 4(a) shows the conjugate vectors as well as their summation vector. In the second column, the real part of the signal and noise vectors are shown. the magnitude of noise alters the angles of the signal conjugate vectors while keeping their magnitude unchanged thus producing conjugate vectors on the circle. It is seen from the column three that the vector produced as a result of adding the modified vectors. Column four demonstrates the real part of the addition vector, while its imaginary part is discarded with a view to avoid getting complex time domain frames after IFFT operation. Comparing column one and four of Fig. 4(a), it is clear that a limited change of original signal occurs if signal vector is greater than noise compensation vector.

In Fig. 4(b), similar illustration is shown by considering signal vector is smaller than noise compensation vector and found that significant change of the original signal occurs. Since the noise compensation vector is anti-symmetric, the angle of the conjugate pair in each case of Fig. 4(b), are pushed in opposite directions, one towards 0 radian and other towards $\pi$ radian. The further they are pushed apart, the more out of phase they become. This justifies that FFT spectrum of noisy speech signal with larger magnitude values undergoes less attenuation and that with smaller magnitude values subject to more attenuation based on the fact that noise frequency components are assumed to have lower magnitude than the clean speech signal, when FFT spectrum of noisy speech has larger magnitude components.

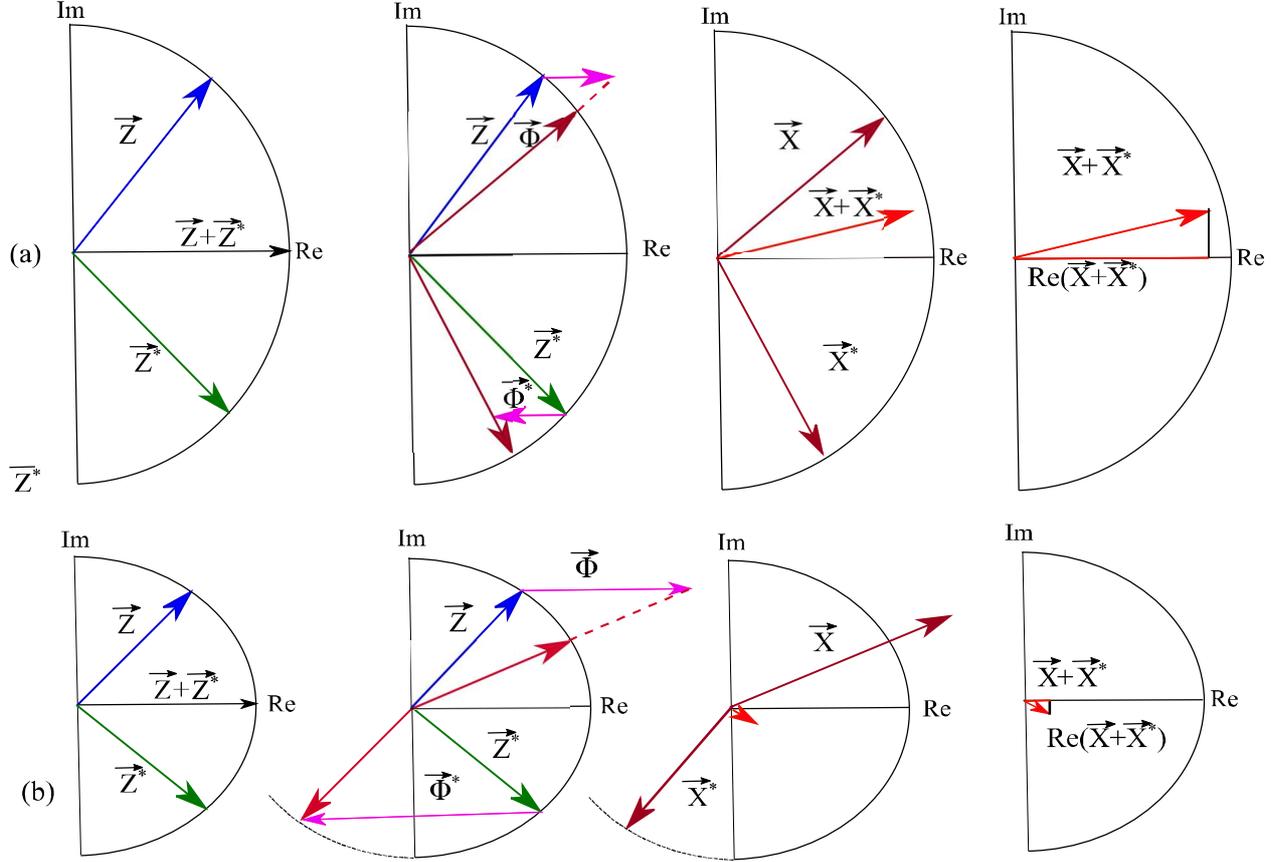

Figure 4: Phase compensation for $\vec{\phi}$ (a) when $|\vec{Z}| > |\vec{\phi}|$ (b) when $|\vec{Z}| < |\vec{\phi}|$.

### 2.4. Resynthesis of enhanced signal

The enhanced speech frame is synthesized by performing the IFFT on the resulting $\widehat{X}^\tau[\omega_k]$,

$$\widehat{x}^\tau[n] = Re\left(IFFT\{\widehat{X}^\tau[\omega_k]\}\right), \tag{20}$$

where $Re(\cdot)$ denotes the real part of the number inside it and $\widehat{x}^\tau[n]$ represents the the enhanced speech frame. The final enhanced speech signal is synthesized by using the standard overlap and add method [29].

### 3. Results

In this Section, a number of simulations is carried out to evaluate the performance of the proposed method.

*3.1. Implementation*

The above proposed method, which we call non-stationary noise-driven geometric approach with phase compensation (NGPC) is implemented in MATLAB R2016b graphical user interface development environment (GUIDE). The MATLAB software with its user manual is attached as supplementary material with the paper. This software also includes implementation of some recent methods such as GA [26], PSC [33] and soft mask estimator with posteriori SNR uncertainty (SMPO) [27]. The implementations of these methods have been taken from publicly available and trusted sources. GA code is taken from http://ecs.utdallas.edu/loizou/cimplants/, PSC implementation code is acquired from http://www.mathworks.com/matlabcentral/fileexchange/30815-phase-spectrum-compensation and SMPO code from http://ecs.utdallas.edu/loizou/cimplants/. The MATLAB implementations of the calculation of segmental and overall SNR improvement are taken from http://ecs.utdallas.edu/loizou/cimplants/ [17].

*3.2. Simulation Conditions*

Real speech sentences from the NOIZEUS database [16] are employed for the evaluations, where the speech data is sampled at 8 KHz. To imitate a noisy environment, noise sequence is added to the clean speech samples at different signal to noise ratio (SNR) levels ranging from −20 dB to 10 dB. Two different types of noise, such as babble and street are adopted from the NOIZEUS database for evaluating the methods both subjectively and objectively. In order to obtain overlapping analysis frames, hamming windowing operation is performed, where size of each of the frame is 96 samples with 50% overlap between successive frames.

*3.3. Comparison Metrics*

Standard Objective metrics [17] namely, segmental SNR (SNRSeg) improvement in dB, overall SNR improvement in dB and perceptual Evaluation of Speech Quality (PESQ) [21] are used for the evaluation of the proposed method. The proposed method is subjectively evaluated in terms of the spectrogram representations. Formal listening tests are also carried out in order to find the analogy between the objective metrics and subjective sound quality. The performance of our method is compared with GA, PSC and SMPO in both objective and subjective senses.

*3.4. Objective Evaluation*

*3.4.1. Results for speech signals with babble noise*

SNRSeg improvement, overall SNR improvement and PESQ scores for speech signals corrupted with multi-talker babble noise for GA, PSC, SMPO and NGPC are shown in Fig. 5, 6 and Table. 1.

In Fig. 5, the performance of the proposed method is compared with those of the other methods at different levels of SNR for babble noise in terms of SNRSeg improvement. We see that the SNRSeg improvement increases as SNR decreases. At a low SNR of −20 dB, the proposed method yields the highest SNRSeg improvement of 4.5 dB, which is significantly higher than GA, PSC and SMPO. Such a large value of SNRSeg improvement at a low level of SNR attest the capability of the proposed method in producing enhanced speech with better quality in adverse environment. At higher SNR levels also, NGPC performs better than other three methods.

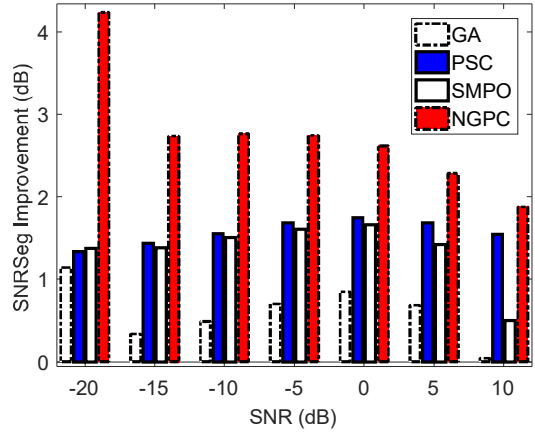

Figure 5: SNRSeg improvement for different methods in babble noise.

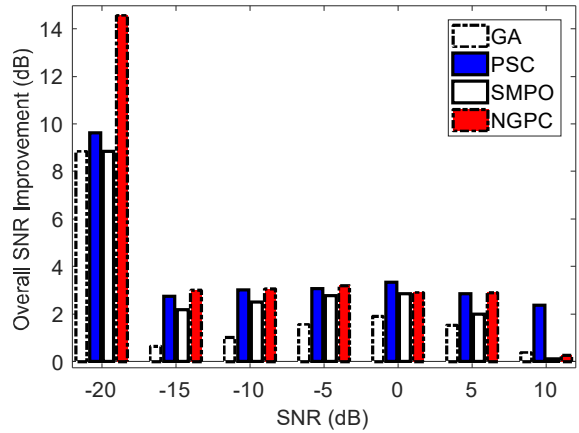

Figure 6: Overall SNR improvement for different methods in babble noise.

Fig. 6 represents the overall SNR improvements as a function of SNR for the proposed method and the other methods. As shown in this figure, NGPC produces a very high overall SNR improvement in −20 dB but shows competitive values at other SNRs in comparison to PSC and SMPO. GA fails completely to show any competitive value.

In Table 1, It can be seen that at a high level of SNR, such as 10 dB, all the methods show better PESQ scores. For the proposed method, PESQ score is competitive and in some cases it is higher than those of other approaches for higher SNRs. But in lower SNRs, the proposed method performs significantly better than other methods in terms of PESQ scores. Since, at a particular SNR, a higher PESQ score indicates a better speech quality, the proposed method is indeed better in performance in the presence of multi-talker babble noise.

Table 1: Comparison of PESQ scores in presence of babble noise

| SNR (in dB) | GA | PSC | SMPO | NGPC |
|---|---|---|---|---|
| -20 | 1.14 | 1.02 | 1.01 | 1.17 |
| -15 | 1.34 | 1.44 | 1.1 | 1.37 |
| -10 | 1.43 | 1.50 | 1.25 | 1.56 |
| -5 | 1.55 | 1.67 | 1.34 | 1.70 |
| 0 | 1.69 | 1.82 | 2.14 | 1.82 |
| 5 | 1.98 | 2.16 | 2.18 | 2.15 |
| 10 | 2.31 | 2.50 | 2.29 | 2.48 |

*3.4.2. Results for Speech Signals with street noise*

SNRSeg Improvement, overall SNR improvement and PESQ scores for speech signals corrupted with street noise for GA, PSC, SMPO and NGPC are shown in Fig. 7, 8 and Table 2.

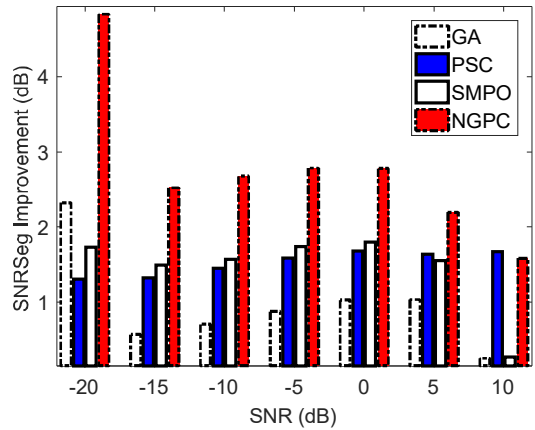

Figure 7: SNRSeg Improvement for different methods in street noise.

In Fig. 7, the performance of the proposed method is compared with those of the other methods at different levels of SNR. We see that the SNRSeg improvement in dB increases as SNR decreases. At a low SNR of −20 dB, NGPC yields the highest SNRSeg improvement score of almost 5.5 dB. For all other SNR levels also, the proposed NGPC method provide significantly higher SNRSeg improvements than other competing methods.

Fig. 8 represents the overall SNR improvement as a function of SNR (in dB) for the proposed method and that for the other methods in the presence of street noise. As shown in the figure, for most of the SNR levels, compared to the other methods, the proposed method is capable of producing enhanced speech with better quality as it gives higher

Table 2: Comparison of PESQ scores in presence of street noise

| SNR (in dB) | GA | PSC | SMPO | NGPC |
|---|---|---|---|---|
| -20 | 1.01 | 1.02 | 1.11 | 1.28 |
| -15 | 1.28 | 1.28 | 1.30 | 1.48 |
| -10 | 1.40 | 1.55 | 1.42 | 1.58 |
| -5 | 1.51 | 1.67 | 1.80 | 1.71 |
| 0 | 1.62 | 1.84 | 1.83 | 1.82 |
| 5 | 1.88 | 2.19 | 2.54 | 2.17 |
| 10 | 2.19 | 2.50 | 2.65 | 2.57 |

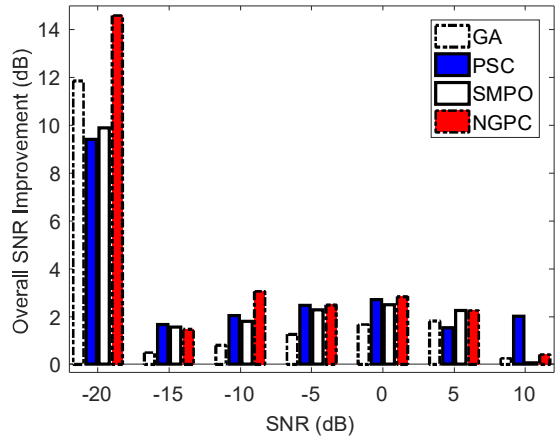

Figure 8: Overall SNR improvement for different methods in street noise.

values of overall SNR improvement.

In Table 2, It can be seen that at a high level of SNR, such as 10 dB, all the methods show higher PESQ scores. But with the decrement of the SNR level, PESQ scores reduce for all the methods. The proposed method provides highre PESQ scores than other methods at most of the lower SNR levels.

3.5. Subjective Evaluation

To evaluate the performance of the proposed method and other competing methods subjectively, we use two commonly used tools. The first one is the plot of the spectrograms of the output for all the methods and compare their performance in terms of preservation of harmonics and capability to remove noise.

The spectrograms of the clean speech, the noisy speech, and the enhanced speech signals obtained by using the

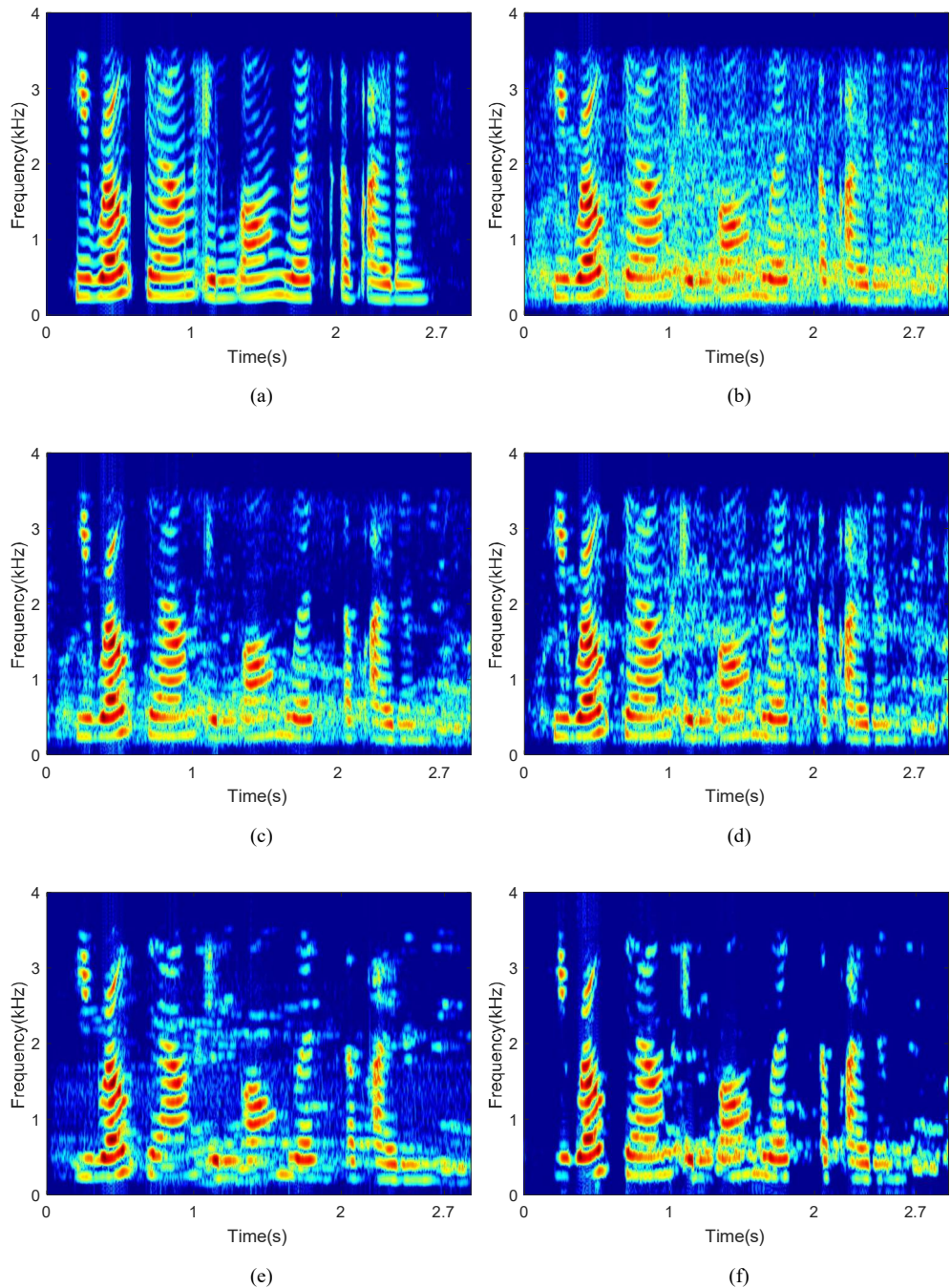

Figure 9: Spectrograms of (a) clean signal (b) noisy signal with 10dB babble noise; spectrograms of enhanced speech from (c) GA (d) PSC (e) SMPO (f) NGPC.

proposed method and all other methods are presented in Fig. 9 for babble noise corrupted speech at an SNR of 10 dB. It is obvious from the spectrograms that the proposed method preserves the harmonics significantly better than all

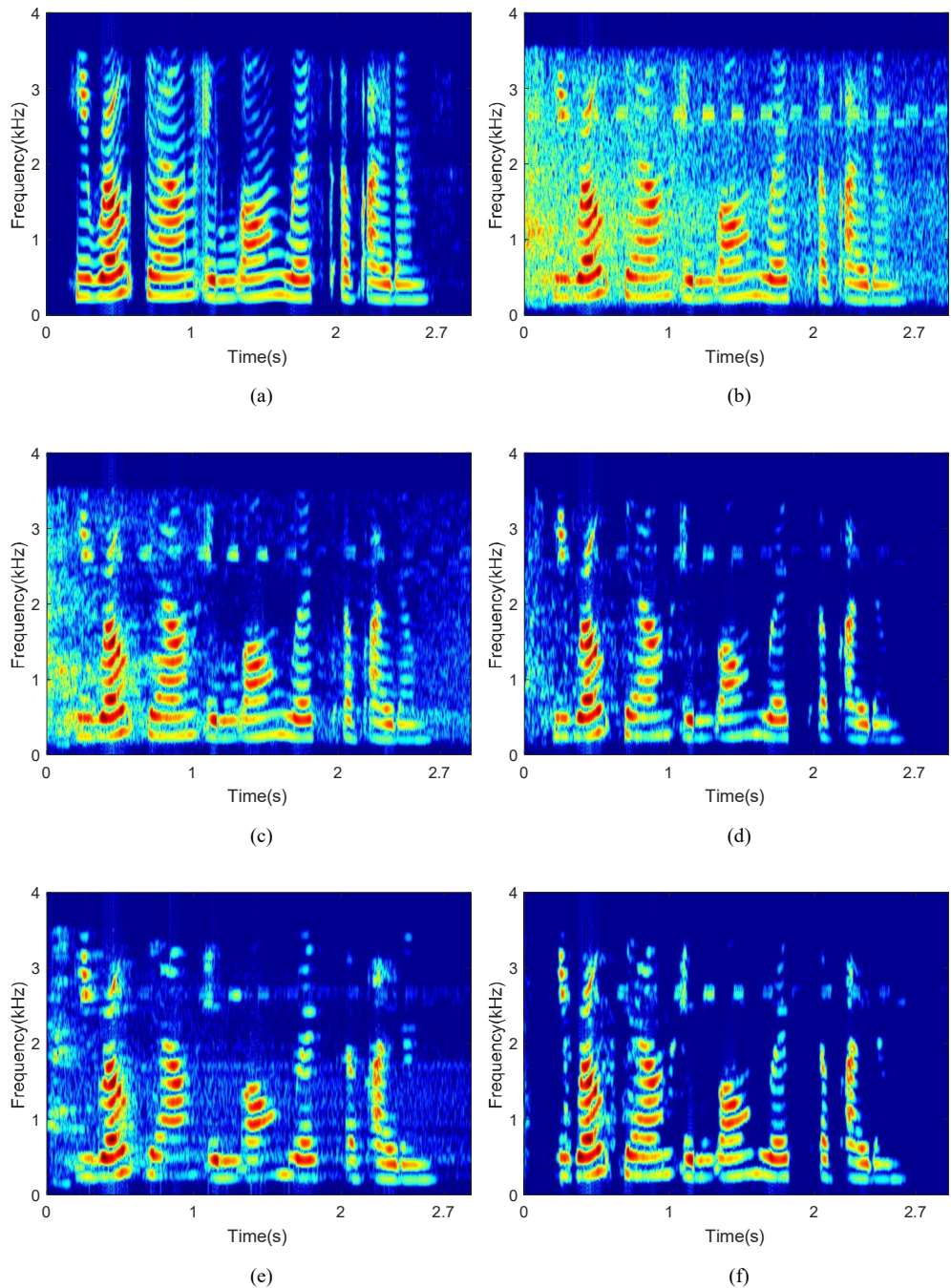

Figure 10: Spectrograms of (a) clean signal (b) noisy signal with 10dB street noise; spectrograms of enhanced speech from (c) GA (d) PSC (e) SMPO (f) NGPC.

the other competing methods. The noise is also reduced at every time point for the proposed method which attest our claim of better performance in terms of higher SNRSeg improvement, higher overall SNR improvement and higher

PESQ values in objective evaluation. Another collection of spectrograms for the proposed method with other methods for speech signals corrupted with street noise is shown in Fig. 10. This figure also attests that our proposed method has better performance in terms of harmonics' preservation and noise removal in presence of street noise.

The second tool we used for subjective evaluation of the proposed method and the competing methods is the formal listening test. We add street and babble noises to all the thirty speech sentences of NOIZEUS database at −20 to 10 SNR levels and process them with all the competing methods. We allow ten listeners to listen to these enhanced speeches from these methods and evaluate them subjectively. Following [20] and [14], We use SIG, BAK and OVL scales on a range of 1 to 5. The detail of these scales and procedure of this listening test is discussed in [20]. More details on this testing methodology of listening test can be obtained from [22].

We show the mean scores of SIG, BAK, and OVRL scales for all the methods for speech signals corrupted with babble noise in Tables 3, 4, and 5 and for speech signals corrupted with street noise is shown in Tables 6, 7, and 8. The higher values for the proposed method in comparison to other methods clearly attest that the proposed method is better than them in terms of lower signal distortion (higher SIG scores), efficient noise removal (higher BAK scores) and overall sound quality (higher OVL scores) for all SNR levels.

Table 3: Mean scores of SIG scale for different methods in presence of babble noise at 5 dB

| Listener | GA | PSC | SMPO | NGPC |
|---|---|---|---|---|
| 1 | 4.1 | 3.6 | 4.0 | 4.4 |
| 2 | 3.8 | 3.3 | 3.9 | 3.5 |
| 3 | 4.1 | 3.9 | 4.0 | 4.6 |
| 4 | 4.1 | 3.4 | 4.2 | 4.5 |
| 5 | 3.8 | 3.2 | 3.8 | 4.5 |
| 6 | 3.7 | 2.9 | 3.6 | 3.8 |
| 7 | 3.5 | 3.8 | 3.8 | 4.5 |
| 8 | 3.6 | 3.4 | 3.6 | 4.8 |
| 9 | 3.6 | 3.5 | 3.9 | 3.8 |
| 10 | 3.7 | 3.7 | 3.8 | 3.8 |

The mean scores in the presence of both street and babble noises demonstrate that lower signal distortion (i.e., higher SIG scores) and lower noise distortion (i.e., higher BAK scores) are obtained with the proposed method relative to other methods in most of the conditions. It is also shown that a consistently better performance in OVRL scale is offered by the proposed method compared to the other methods. Thus, we conclude that the proposed method possesses the highest subjective sound quality compared to the other methods in case of different noises at various levels of SNR.

Table 4: Mean scores of BAK scale for different methods in presence of babble noise at 5 dB

| Listener | GA | PSC | SMPO | NGPC |
|---|---|---|---|---|
| 1 | 3.5 | 4.0 | 4.5 | 4.7 |
| 2 | 4.2 | 4.3 | 4.9 | 4.4 |
| 3 | 4.3 | 4.2 | 4.4 | 4.9 |
| 4 | 3.7 | 4.4 | 4.7 | 4.5 |
| 5 | 4.6 | 4.2 | 4.8 | 4.7 |
| 6 | 4.5 | 3.9 | 4.6 | 4.6 |
| 7 | 3.8 | 3.8 | 3.9 | 4.7 |
| 8 | 4.7 | 4.4 | 4.6 | 4.6 |
| 9 | 3.8 | 3.5 | 3.9 | 4.6 |
| 10 | 4.7 | 4.7 | 4.8 | 4.5 |

## 4. Conclusions

An improved approach to solve the problem of speech enhancement using the geometric approach of spectral subtraction and phase spectrum compensation has been presented in this article. Using these two methods, we compensate the spectrum of noisy speech in two steps to get an enhanced speech signal. The proposed method performs better than conventional methods which traditionally adopt only either of the magnitude compensation or phase compensa-tion. Simulation results show that the proposed method yields consistently better results in terms of higher SNRSeg Improvements, higher PESQ values, and higher overall SNR improvements than the existing speech enhancement methods. A subjective listening test over a broad range of noise types and SNR levels among a large number of listeners has also supported the efficacy of our proposed method in producing a better enhanced speech.

Table 5: Mean scores of OVL scale for different methods in presence of babble noise at 5 dB

| Listener | GA | PSC | SMPO | NGPC |
|---|---|---|---|---|
| 1 | 2.3 | 2.9 | 4.0 | 4.5 |
| 2 | 2.3 | 3.4 | 3.8 | 4.4 |
| 3 | 2.4 | 3.4 | 4.1 | 4.5 |
| 4 | 2.3 | 3.3 | 4.2 | 4.2 |
| 5 | 2.3 | 3.2 | 3.9 | 4.5 |
| 6 | 2.2 | 3.7 | 4.6 | 4.2 |
| 7 | 2.3 | 3.4 | 3.8 | 4.4 |
| 8 | 2.3 | 3.6 | 4.1 | 4.5 |
| 9 | 2.3 | 3.1 | 4.5 | 4.4 |
| 10 | 2.4 | 3.1 | 4.8 | 4.4 |

Table 6: Mean score of SIG scale for different methods in presence of street noise at 5 dB

| Listener | GA | PSC | SMPO | NGPC |
|---|---|---|---|---|
| 1 | 3.6 | 4.0 | 4.1 | 4.0 |
| 2 | 3.7 | 4.0 | 4.2 | 4.1 |
| 3 | 3.5 | 4.1 | 3.9 | 4.1 |
| 4 | 3.5 | 4.1 | 3.8 | 3.9 |
| 5 | 3.6 | 3.9 | 3.7 | 4.0 |
| 6 | 3.5 | 4.1 | 4.2 | 4.2 |
| 7 | 3.1 | 4.2 | 4.2 | 3.9 |
| 8 | 3.5 | 4.0 | 3.7 | 4.0 |
| 9 | 3.5 | 4.0 | 3.9 | 4.1 |
| 10 | 3.5 | 4.1 | 4.1 | 4.0 |

Table 7: Mean scores of BAK scale for different methods in presence of street noise at 5 dB

| Listener | GA | PSC | SMPO | NGPC |
|---|---|---|---|---|
| 1 | 2.3 | 2.9 | 4.1 | 4.5 |
| 2 | 2.3 | 3.5 | 3.7 | 4.4 |
| 3 | 2.4 | 3.5 | 4.2 | 4.5 |
| 4 | 2.3 | 3.4 | 4.1 | 4.2 |
| 5 | 2.3 | 3.3 | 3.8 | 4.5 |
| 6 | 2.2 | 3.6 | 4.7 | 4.2 |
| 7 | 2.3 | 3.5 | 3.9 | 4.4 |
| 8 | 2.3 | 3.6 | 4.2 | 4.5 |
| 9 | 2.3 | 3.2 | 4.6 | 4.4 |
| 10 | 2.4 | 3.3 | 4.7 | 4.4 |

Table 8: Mean scores of OVL scale for different methods in presence of street noise at 5 dB

| Listener | GA | PSC | SMPO | NGPC |
|---|---|---|---|---|
| 1 | 2.8 | 4.0 | 4.1 | 4.2 |
| 2 | 2.9 | 4.1 | 3.9 | 4.1 |
| 3 | 2.8 | 4.0 | 4.0 | 4.3 |
| 4 | 2.7 | 4.1 | 4.2 | 4.2 |
| 5 | 2.9 | 4.0 | 4.3 | 4.3 |
| 6 | 2.8 | 4.1 | 4.1 | 4.2 |
| 7 | 2.7 | 4.1 | 3.9 | 4.2 |
| 8 | 2.8 | 4.0 | 4.1 | 4.3 |
| 9 | 2.8 | 4.2 | 4.1 | 4.2 |
| 10 | 2.8 | 4.0 | 4.2 | 4.3 |